\documentclass[conference,compsoc]{IEEEtran}
\IEEEoverridecommandlockouts
\usepackage{cite}
\usepackage{amsmath,amssymb,amsfonts}
\usepackage{textcomp}
\usepackage[utf8]{inputenc} 
\usepackage[T1]{fontenc}    
\usepackage{hyperref}       
\usepackage{url}            
\usepackage{booktabs}       
\usepackage{amsfonts}       
\usepackage{nicefrac}       
\usepackage{microtype}      
\usepackage{xcolor}         
\usepackage{graphicx}
\usepackage{amsmath} 
\usepackage{multirow} 
\usepackage{cleveref}
\usepackage{enumitem}
\usepackage{pgfplots}
\usepackage{tikz}
\usepackage{algorithmic}
\usepackage{pifont}
\usepackage{subcaption}
\usepackage{xspace}
\usepackage{tcolorbox}
\usepackage[linesnumbered,ruled,vlined]{algorithm2e}
\usepackage{listings}
\def\BibTeX{{\rm B\kern-.05em{\sc i\kern-.025em b}\kern-.08em
    T\kern-.1667em\lower.7ex\hbox{E}\kern-.125emX}}

\newcommand{\sysname}{{\sc ParVAL}\xspace}

\begin{document}

\title{Large Language Models for Validating Network Protocol Parsers}

\author{\IEEEauthorblockN{Mingwei Zheng}
\IEEEauthorblockA{\textit{Department of Computer Science} \\
\textit{Purdue University}\\
West Lafayette, USA \\
zheng618@purdue.edu}
\and
\IEEEauthorblockN{Danning Xie}
\IEEEauthorblockA{\textit{Department of Computer Science} \\
\textit{Purdue University}\\
West Lafayette, USA \\
xie342@purdue.edu}
\and
\IEEEauthorblockN{Xiangyu Zhang}
\IEEEauthorblockA{\textit{Department of Computer Science} \\
\textit{Purdue University}\\
West Lafayette, USA \\
xyzhang@purdue.edu}
}

\maketitle
\newcommand{\q}[1]{``{#1}''}

\newcommand{\code}[1]{\texttt{\small#1}\xspace}

\newcommand{\todoc}[2]{{\textcolor{#1}{\textbf{#2}}}}
\newcommand{\todoblack}[1]{{\todoc{black}{\textbf{[[#1]]}}}}
\newcommand{\todored}[1]{{\todoc{red}{\textbf{[[#1]]}}}}
\definecolor{applegreen}{rgb}{0.55, 0.71, 0.0} 
\newcommand{\todogreen}[1]{\todoc{applegreen}{\textbf{[[#1]]}}}
\newcommand{\todoblue}[1]{\todoc{blue}{\textbf{[[#1]]}}}
\newcommand{\todoorange}[1]{\todoc{orange}{\textbf{[[#1]]}}}
\newcommand{\todobrown}[1]{\todoc{brown}{\textbf{[[#1]]}}}
\newcommand{\todogray}[1]{\todoc{gray}{\textbf{[[#1]]}}}
\newcommand{\todopurple}[1]{\todoc{purple}{\textbf{[#1]}}}
\newcommand{\todopink}[1]{\todoc{magenta}{\textbf{[[#1]]}}}
\newcommand{\todocyan}[1]{\todoc{cyan}{\textbf{[[#1]]}}}
\newcommand{\todoviolet}[1]{\todoc{violet}{\textbf{[[#1]]}}}
\newcommand{\todoteal}[1]{\todoc{teal}{\textbf{[[#1]]}}}
\newcommand{\todo}[1]{\todored{TODO: #1}}

\definecolor{light-gray}{gray}{0.7}
\newcommand{\hilight}[1]{\colorbox{light-gray}{#1}}

\makeatletter
\newcommand*{\textoverline}[1]{$\overline{\hbox{#1}}\m@th$}
\makeatother

\newcommand\blackcircle[1]{%
  \tikz[baseline=(X.base)] 
    \node (X) [draw, shape=circle, inner sep=0, scale=0.8, fill=darkgray, text=white] {\strut #1};%
}

\newcommand\whitecircle[1]{%
  \tikz[baseline=(X.base)] 
    \node (X) [draw, shape=circle, inner sep=0, scale=0.8, fill=white, text=black] {\strut #1};%
}



\newif\ifenablecomments  
\enablecommentstrue      

\ifenablecomments
    \newcommand{\mingwei}[1]{\todoblue{Mingwei: #1}}
    \newcommand{\danning}[1]{\todogreen{Danning: #1}}
    \newcommand{\wang}[1]{\todobrown{wang: #1}}
    \newcommand{\xz}[1]{\todored{XZ: #1}}
\else
    \newcommand{\mingwei}[1]{}
    \newcommand{\danning}[1]{}
    \newcommand{\wang}[1]{}
    \newcommand{\xz}[1]{}
\fi

\newboolean{showchanges}
\setboolean{showchanges}{true} 

\newcommand{\change}[1]{%
    \ifthenelse{\boolean{showchanges}}%
        {\textcolor{blue}{#1}}
        {#1}
}


\newcounter{finding}
\newcommand{\intuition}[1]{
\begin{tcolorbox}[colback=gray!10, colframe=gray!20, boxrule=0pt, boxsep=2pt,
left=2pt, right=2pt, top=1pt, bottom=1pt, sharp corners, leftrule=2mm, leftrule=2pt, colframe=gray]
\refstepcounter{finding}
\textbf{Conclusion~\thefinding{}:} \emph{#1}
\end{tcolorbox}
}

\newcommand{\tool}{{[toolName]}\xspace}

\captionsetup[figure]{font=bf,skip=6pt}
\captionsetup[table]{font=bf,skip=6pt}
\newcommand{\distance}{8pt}
\setlength{\textfloatsep}{6pt}
\setlength{\floatsep}{\distance}
\setlength{\intextsep}{\distance}
\setlength{\dbltextfloatsep}{\distance} 
\setlength{\dblfloatsep}{\distance} 

\begin{abstract}
Network protocol parsers are essential for enabling correct and secure communication between devices. Bugs in these parsers can introduce critical vulnerabilities, including memory corruption, information leakage, and denial-of-service attacks.
An intuitive way to assess parser correctness is to compare the implementation with its official protocol standard.
However, this comparison is challenging because protocol standards are typically written in natural language, whereas implementations are in source code. Existing methods like model checking, fuzzing, and differential testing have been used to find parsing bugs, but they either require significant manual effort or ignore the protocol standards, limiting their ability to detect semantic violations.
To enable more automated validation of parser implementations against protocol standards, we propose \sysname, a multi-agent framework built on large language models (LLMs).
\sysname leverages the capabilities of LLMs to understand both natural language and code. It transforms both protocol standards and their implementations into a unified intermediate representation, referred to as format specifications, and performs a differential comparison to uncover inconsistencies.
We evaluate \sysname on the Bidirectional Forwarding Detection (BFD) protocol. Our experiments demonstrate that \sysname successfully identifies inconsistencies between the implementation and its RFC standard, achieving a low false positive rate of 5.6\%. \sysname uncovers seven unique bugs, including five previously unknown issues.
\end{abstract}


\section{Introduction}
Network protocols specify the rules that control communication between devices, for example, TLS~\cite{TLS} for secure communication, FTP~\cite{FTP} for file transfers, and IP~\cite{IP} for routing data. 
Network protocol parsing is a critical component of protocol implementations, responsible for analyzing incoming packets, validating their correctness, and discarding malformed ones.
This process protects data integrity and enhances system security by preventing the processing of potentially harmful packets.

Despite their critical role, network protocol parsers are error-prone, which can lead to severe security vulnerabilities, including memory corruption, information leakage, and denial-of-service attacks. 
For instance, the Heartbleed~\cite{heartbleed} vulnerability in OpenSSL~\cite{OpenSSL}’s TLS parser allowed attackers to access sensitive user information due to a missing bounds check. Similarly, CVE-2021-41773~\cite{CVE-2021-41773} in Apache’s HTTP server exposed confidential files through a path traversal flaw in the request parser.

Validating the correctness of network protocol parsers is inherently challenging. A natural approach is to compare the parser implementation against its official protocol standard to identify inconsistencies.
However, this process is complicated by a fundamental representational gap: protocol standards are typically written in natural language (e.g., RFCs), while implementations are written in source code.
Although both aim to specify the same behavior, their different representations hinder direct comparison, making systematic validation a significant challenge.

Various techniques have been proposed to address this challenge.
Model checking~\cite{MusuvathiE04, DiazCRP04, Magic} verifies parser behavior against formal protocol models to detect logical errors. While effective, it often requires manual construction of formal models, making it impractical for applying to large and frequently evolving protocols.
Fuzzing~\cite{pham2020aflnet, shi2023lifting} automatically produces large volumes of test inputs to expose memory-related issues, such as crashes.
However, it primarily targets low-level bugs and fails to assess whether the parser conforms to the protocol's intended semantics, leaving many semantic bugs~\cite{Pardiff, KIT, 9678600, ParCleanse} undetected.
Differential analysis~\cite{Pardiff, reen2020dpifuzz} identifies inconsistencies by comparing how different implementations handle the same input.
This approach assumes that at least one implementation behaves correctly and may miss bugs shared across all versions.
While these methods are effective in certain situations, they fall short of providing an efficient way to validate parser implementations against official protocol standards written in natural language.

\vspace{1mm}
\noindent\textbf{Our Approach.} We propose \sysname, an LLM-based multi-agent framework for validating parser implementations against their official protocol standards.
\sysname uses LLMs to automatically abstract both the natural language documentation and the parser implementation into a unified intermediate representation, called format specifications.
These specifications capture the expected packet structure and parsing logic, enabling direct comparison between the protocol documentation and its implementation.

LLMs have demonstrated strong capabilities in extracting format specifications from natural language~\cite{fakhoury20243dgen, ParCleanse}, but applying them to source code remains challenging. In real-world codebases, parsing logic is often scattered across multiple files and functions~\cite{Pardiff, shi2023lifting}, making it difficult to gather the full context needed for accurate format extraction.
Moreover, LLMs are prone to generating incorrect or incomplete outputs (hallucinations), so validating and refining the extracted formats is essential. This typically relies on unit tests to check whether the extracted format matches the implementation behavior. However, without a well-isolated parsing module, generating and executing such tests is difficult.
Consequently, errors in the extracted format specifications may go unnoticed, undermining the reliability of comparing the format specifications extracted from the parser implementation and the protocol standard.

To tackle the challenge of retrieving parsing-relevant context from large codebases,  \sysname introduces a \textit{Retrieval-Augmented Program Analysis Agent} to automatically extract code segments directly related to protocol parsing while excluding irrelevant ones.
This agent uses a \textit{dependency-aware retrieval mechanism}, beginning from the entry parsing function and recursively using LLMs to analyze data dependencies, control flow, and function calls. It incrementally collects parsing relevant segments until the complete parsing logic is isolated.

To reduce LLM hallucinations in LLM-generated format specifications, \sysname isolates the extracted parsing logic into a standalone, executable module.
This module takes a buffer and its length as input and returns a boolean value indicating whether parsing succeeds. 
\sysname then extracts a format specification from this module and generates unit tests to check whether the module's behavior matches the extracted format specification.
Any discrepancies identified through testing are used as feedback to iteratively refine the format, improving its accuracy over time.

\vspace{1mm}
\noindent\textbf{Contributions.}
This paper explores the potential of LLMs to validate parser implementations against natural language protocol standards. 
While our approach does not guarantee soundness or completeness, our evaluation on a real-world protocol demonstrates its practical effectiveness, providing a strong starting point for further investigation.
In summary, our main contributions are:

\begin{itemize}[leftmargin=0.5cm]
    \item We introduce \sysname, a multi-agent framework that uses LLMs to extract parsing behaviors from both source code and protocol standards. These behaviors are represented in a unified intermediate representation, called format specifications, to enable direct comparison.
    \item \sysname introduces a \textit{dependency-aware retrieval mechanism} that automatically identifies and extracts relevant parsing logic by analyzing data flow, control flow, and function dependencies throughout the codebase.
    \item \sysname isolates parsing logic into a standalone, executable module, enabling test case generation to validate and refine the extracted format specifications.
   This helps improve the accuracy of format specification extraction and mitigates the impact of LLM hallucination.
    We compare the extracted format specification against a manually annotated ground truth and find that \sysname successfully extract the full parsing logic.
 
    \item We implement the proposed approach in \sysname\footnote{\sysname is available at: \url{https://github.com/zmw12306/PARVAL}}.
    We have evaluated \sysname on the Bidirectional Forwarding Detection (BFD) protocol to validate its parser implementation against the corresponding RFC standard. \sysname successfully identifies seven unique bugs, including five previously undiscovered.
    
\end{itemize}

\noindent\textbf{Organization.}
\Cref{sec:motiv} presents a motivation example to illustrate the challenges and our solutions.
\Cref{sec:approach} details our design.
\Cref{sec:eval} evaluates the effectiveness of our tool.
\Cref{sec:discussion} discusses limitations and future work.
\Cref{sec:related} reviews related work.
\Cref{sec:conclusion} concludes the paper.

\section{Motivating Example}
\label{sec:motiv}

\begin{figure}[t]
    \centering
    \includegraphics[width=\linewidth]{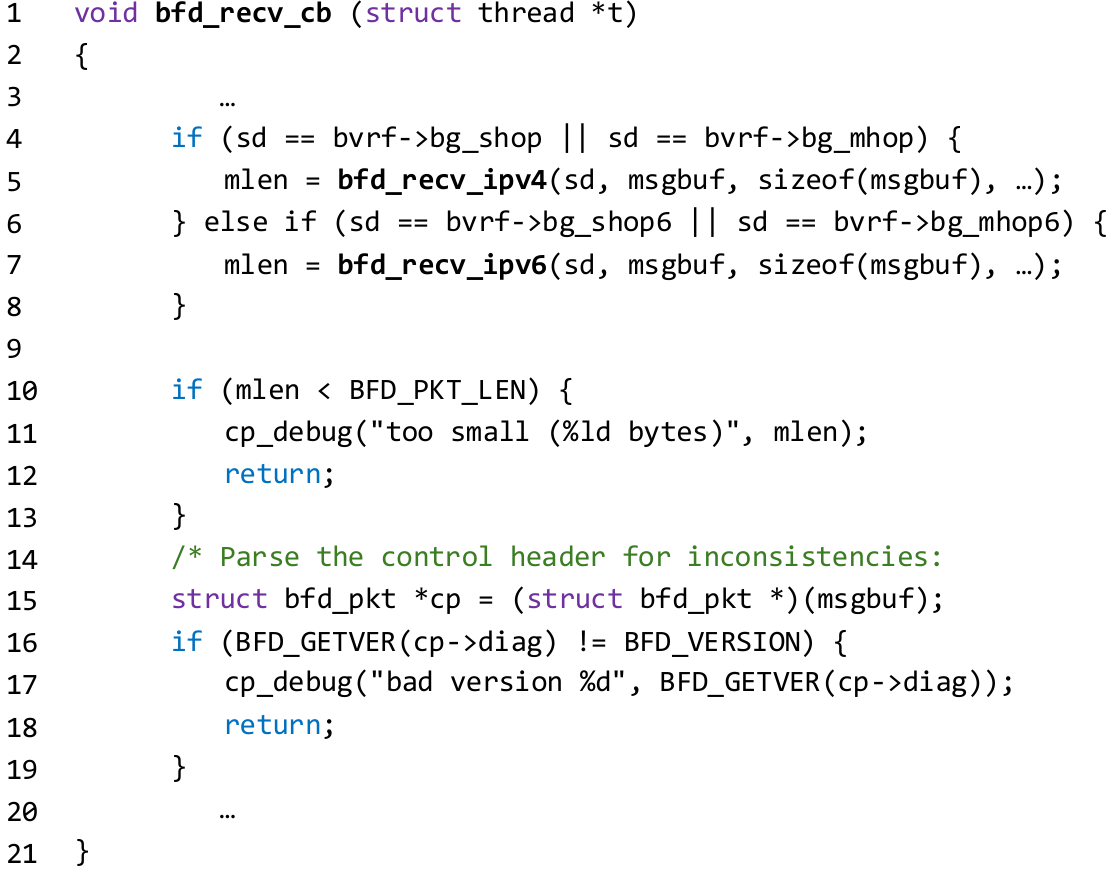}
    \caption{Entry Parsing Function in BFD Implementation}
    \label{fig:entryparser}
    
\end{figure}

\Cref{fig:entryparser}
shows a simplified version of the entry parsing function in the BFD implementation. 
The function \texttt{bfd\_recv\_cb} handles incoming BFD packets. 
It begins by determining the socket type (\texttt{sd}) and invokes either \texttt{bfd\_recv\_ipv4} or \texttt{bfd\_recv\_ipv6} to read the message into \texttt{msgbuf} (lines 4 - 8). 
It then validates the message length \texttt{mlen} (lines 10 - 13) and proceeds to parse the control header (lines 14 - 20). 

To validate the parser implementation, we leverage large language models (LLMs) to extract format specifications from the source code and the RFC document.
By translating both into a unified representation, we enable direct comparison and systematic validation. However, accurately extracting protocol formats from source code presents two key challenges.

\vspace{1mm}
\noindent\textbf{Challenge 1: Retrieving Parsing-Related Context.} 
LLMs require complete parsing contexts to accurately extract protocol formats. However, in real-world codebases, parsing-relevant logic is often spread across multiple functions, files, and directories, making it challenging to automatically identify and extract all relevant code.
For example, in \Cref{fig:entryparser}, \texttt{struct bfd\_pkt} (line 15) and \texttt{BFD\_GETVER} (line 16) are defined in separate files.
Without retrieving these dependencies, LLMs cannot fully interpret the parsing logic. Since feeding the entire codebase exceeds the context window of LLMs, a selective context retrieval strategy is necessary.

\vspace{1mm}
\noindent\textbf{Our Solution}: We design a \textit{dependency-aware retrieval mechanism} that guides LLMs to selectively collect parsing-relevant code. Starting from the entry parsing function, it traces input buffer usage through data and control dependencies, as well as function calls, to collect essential context such as structs, macros, and helper functions. This dependency analysis guides LLMs to retrieve only the necessary context to accurately extract protocol formats.

\vspace{2mm}
\noindent\textbf{Challenge 2: LLM Hallucinations in Extracted Formats.}
LLMs may generate inaccurate format specifications that deviate from actual parser behavior, such as involving incorrect field types, lengths, or ordering.
To correct these hallucinations, it is essential to test the consistency between the extracted format specification and the original parser, using the test results to guide iterative refinement.
However, this feedback loop becomes unreliable when the parsing logic is tightly coupled with other parsing unrelated operations.
For instance, in \Cref{fig:entryparser}, the function \texttt{bfd\_recv\_cb} expects a \texttt{thread} structure containing socket state and runtime configuration. Before any parsing occurs, the function performs socket-related checks.
In contrast, test inputs generated from the extracted format consist solely of raw packet data and cannot be passed directly to \texttt{bfd\_recv\_cb}. 
To run a test, each packet data must be wrapped in a correctly initialized \texttt{thread} structure. This additional setup introduces complexity and obscures whether test failures stem from the format itself, incorrect \texttt{thread} initialization, or buggy socket handling.
This undermines the reliability of test-based feedback, making it difficult to detect and correct LLM hallucinations in the extracted format.

\vspace{1mm}
\noindent\textbf{Our Solution}: To reduce LLM hallucinations, \sysname extracts the parsing logic into a standalone, executable module that focuses only on packet parsing. This module takes a raw buffer and its length as input and returns a boolean indicating whether parsing succeeds.
Valid and invalid test packets generated from the extracted format specification are directly passed to this module.
If the module rejects valid packets or accepts invalid ones, this indicates a mismatch between the LLM extracted format specification and the actual parsing behavior.
Such mismatches are used to iteratively refine the extracted format, improving its accuracy.

\section{Approach}
\begin{figure}[t]
	\centering
	\includegraphics[width=\linewidth]{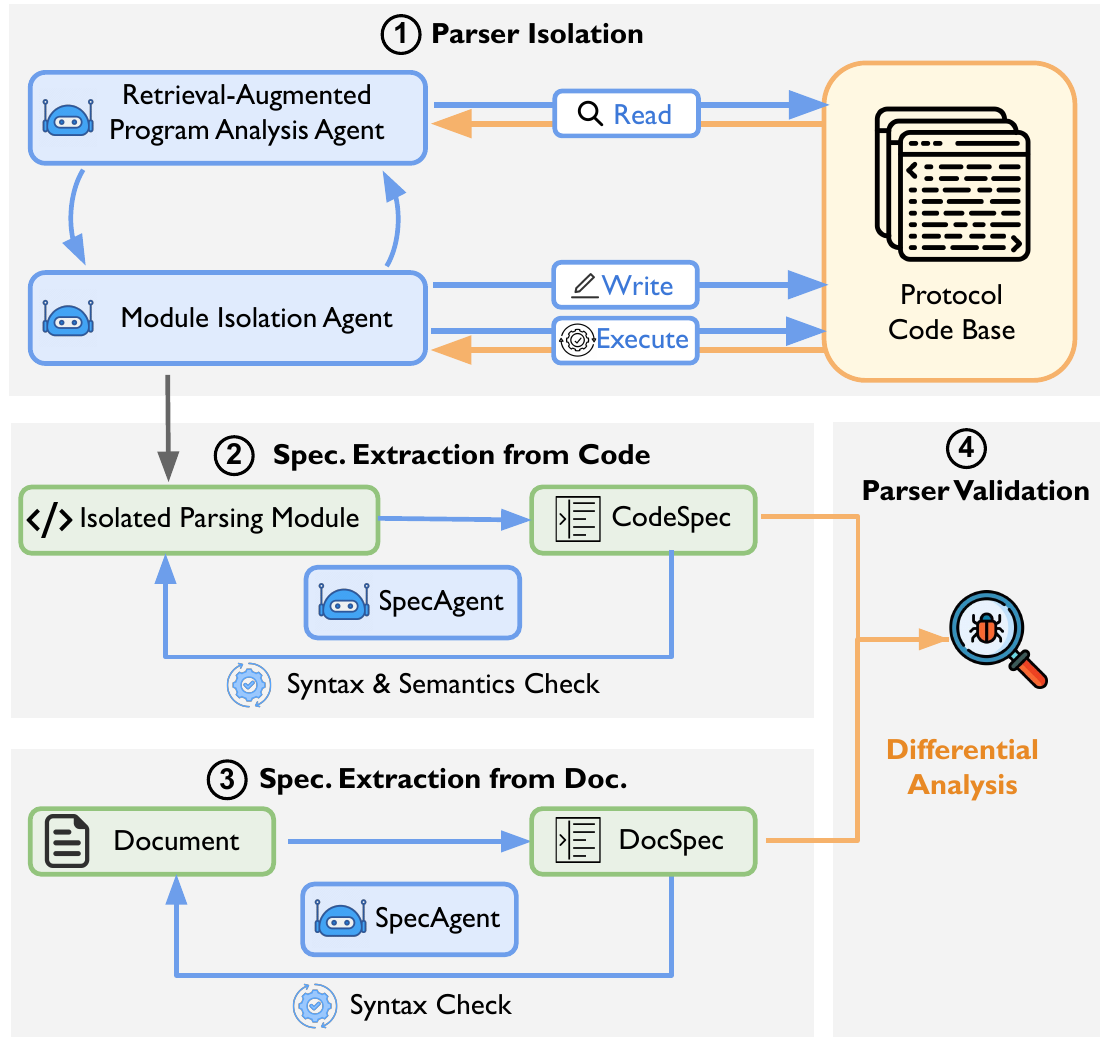}
	\caption{The Overview of \sysname}
	\label{fig:overview}
\end{figure}

\Cref{fig:overview} presents an overview of our approach, which consists of four stages. In \textbf{Stage 1 Parser Isolation} (\Cref{subsec:stage1}), the \textit{Retrieval-Augmented Program Analysis Agent} and the \textit{Module Isolation Agent} collaborate to extract an \textit{Isolated Parsing Module} from the protocol codebase. In \textbf{Stage 2 Spec. Extraction from Code} (\Cref{subsec:stage2}) and \textbf{Stage 3 Spec. Extraction from Doc.}~(\Cref{subsec:stage3}), the \textit{Spec Agent} extracts format specifications from the \textit{Isolated Parsing Module} and protocol standards, producing \textit{CodeSpec} and \textit{DocSpec}, respectively.
In \textbf{Stage 4 Parser Validation} (\Cref{subsec:stage4}), the two format specifications are compared to identify inconsistencies, which may point to bugs in the implementation or missing details in the protocol standard.
The following sections describe each stage in detail.

\label{sec:approach}

\label{sec:approach}
\begin{figure}[t]
	\centering
	\includegraphics[width=\linewidth]{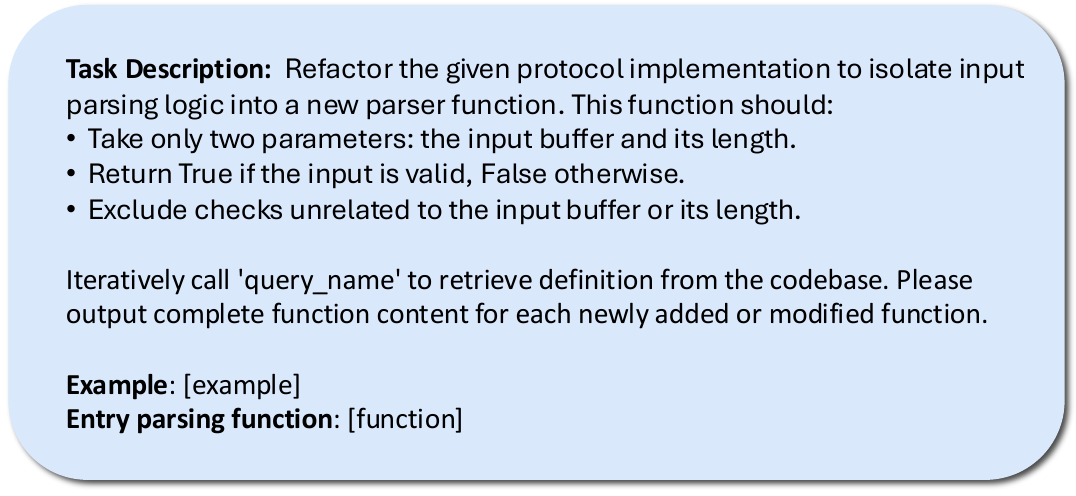}
	\caption{LLM Task Description for Parser Isolation}
	\label{fig:isolate}
\end{figure}
\subsection{Stage 1: Parser Isolation}\label{subsec:stage1}
In this stage, we isolate the parsing logic from the protocol codebase using an LLM-based multi-agent framework that integrates \textit{dependency-aware retrieval mechanism} and \textit{parser module construction}. As shown in \Cref{fig:overview}, the framework consists of two LLM agents: the \textit{Retrieval-augmented Program Analysis Agent} and the \textit{Module Isolation Agent}, augmented by tools to interact (i.e., read, write, or execute) with the protocol codebase. The task description for this multi-agent system is shown in \Cref{fig:isolate}.

The \textbf{\textit{Retrieval-augmented Program Analysis Agent}} uses an LLM to iteratively analyze code and retrieve relevant context through a \textbf{\textit{dependency-aware retrieval mechanism}}. Starting from the entry parsing function, it traces input buffer usage via data and control dependencies, as well as function calls, to identify macros, types, and helper functions critical to parsing.
It then employs an AST-based retrieval tool, built on Tree-sitter~\cite{treesitter}, to extract the corresponding definitions from the source code (\textbf{read}).
In parallel, the \textbf{\textit{Module Isolation Agent}} utilizes the retrieved context to construct an \textit{isolated parsing module}. 
It iteratively \textbf{writes} identified parsing logic into a standalone, executable module and \textbf{executes} it, continuously refining the module based on build results and error feedback.
When additional definitions or clarifications are needed, it collaborates with the \textit{Retrieval-augmented Program Analysis Agent} to retrieve more contexts. 
The final output is a fully \textbf{\textit{isolated parsing module}} that takes an input buffer and its length as input and returns a boolean indicating parsing success.
Unlike the original codebase, which often mixes parsing with other processing logic, this isolated module contains only the parsing logic and is independently testable.

\vspace{1mm}

\begin{figure}[t]
    \centering
    \includegraphics[width=\linewidth]{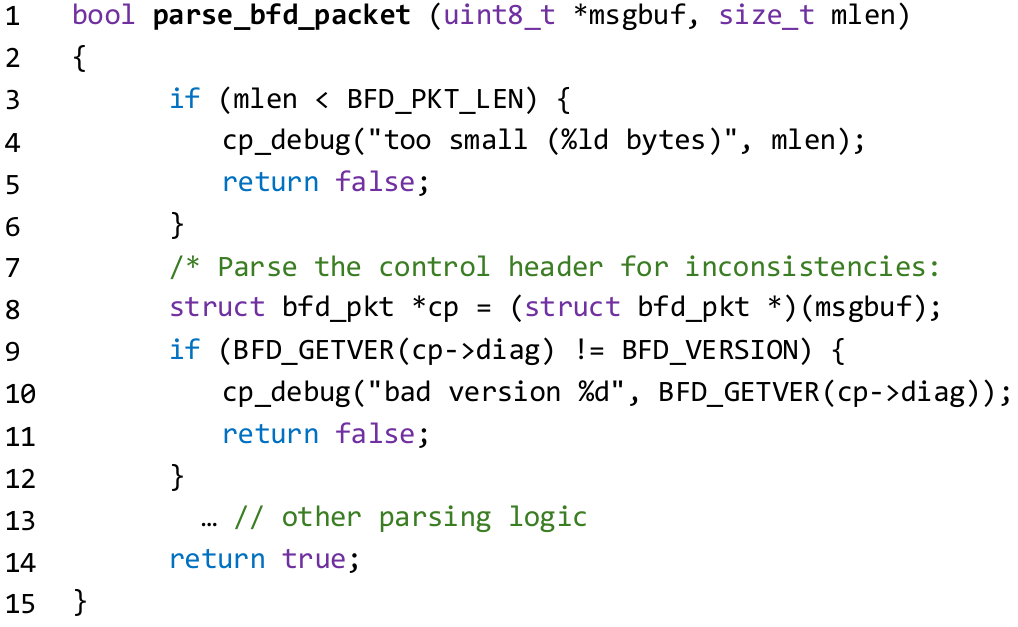}
    \caption{\textit{Isolated Parsing Module} Example}
    \label{fig:isolatedparser}
\end{figure}

\noindent\textbf{Example.} For the entry parsing function shown in \Cref{fig:entryparser}, isolating the parsing logic requires understanding the behavior of its callee functions, as additional parsing logic may reside within them.  
To achieve this, LLMs must retrieve relevant context, just as a human would when analyzing the code.
Hence, the \textit{Retrieval-augmented Program Analysis Agent} retrieves the definition of \texttt{bfd\_recv\_ipv4}, \texttt{bfd\_recv\_ipv6}, and \texttt{BFD\_GETVER} to determine whether they contain parsing-related operations. 
The agent identifies \texttt{msgbuf} and \texttt{mlen} as the received BFD packet and its length.
Therefore, determining whether a piece of code contributes to BFD parsing involves analyzing its dependencies on \texttt{msgbuf} and \texttt{mlen}.
For example, at line 15 in \Cref{fig:entryparser}, \texttt{msgbuf} is cast into a \texttt{bfd\_pkt}, making \texttt{cp} directly data-dependent on \texttt{msgbuf}. The \texttt{if} condition at line 16 operates on a field of \texttt{cp} and performs input validation, making it directly relevant to BFD packet parsing.
In contrast, the \texttt{if} conditions at lines 4 and 6 check the socket types rather than BFD packet fields, hence irrelevant to BFD packet parsing.
The resulting \textit{Isolated Parsing Module} generated by the \textit{Module Isolation Agent} is shown in \Cref{fig:isolatedparser}. 

\subsection{Stage 2: Specification Extraction from Code}\label{subsec:stage2}
In this stage, we employ \textbf{\textit{SpecAgent}} to transform the \textit{isolated parsing module} into a formal representation of the input format. This representation, called \textbf{\textit{CodeSpec}}, is written in a domain-specific language (DSL) for describing protocol formats.
\textit{SpecAgent} performs this translation automatically by analyzing the parsing logic and iteratively refining the resulting specification based on feedback from both syntax and semantic checks.

\vspace{1mm}
\noindent\textbf{Syntax Check.}
To ensure syntactic correctness, \textit{SpecAgent} uses the DSL's grammar to generate an initial \textit{CodeSpec} from the \textit{isolated parsing module}. It then iteratively validates and refines the specification until it conforms to the DSL rules.

\vspace{1mm}
\noindent\textbf{Semantic Check.}
Beyond syntactic correctness, \textit{CodeSpec} must match the semantic behavior of the \textit{isolated parsing module}. We verify this by generating symbolic test cases from \textit{CodeSpec}—positive cases that conform to the format and negative ones that violate it. These are run against the \textit{isolated parsing module} to check whether valid inputs are accepted and malformed ones are rejected. If a positive case fails or a negative case passes, it indicates a semantic mismatch. 
To resolve such discrepancies, \textit{SpecAgent} iteratively refines \textit{CodeSpec} based on the test outcomes. Specifically, when a positive case fails, we instrument the parser to capture its execution trace, which is then provided to \textit{SpecAgent} to guide the correction. In contrast, if a negative case is incorrectly accepted, \textit{SpecAgent} analyzes how the input violates \textit{CodeSpec} and adjusts the format to enforce the intended parsing behavior.

\subsection{Stage 3: Specification Extraction from Doc.}\label{subsec:stage3}
In this stage, \textbf{\textit{SpecAgent}} analyzes protocol standards (e.g., RFCs) to extract a structured format specification, \textbf{\textit{DocSpec}}, using the same DSL as in Stage 2. This transforms the natural language description of the message format—such as field definitions, length constraints, and value ranges—into a formal representation. Using a unified DSL enables direct, systematic comparison between \textit{DocSpec} and the code-derived \textit{CodeSpec}.

\subsection{Stage 4: Parser Validation}\label{subsec:stage4}

In the final stage, we validate the parser implementation by comparing the protocol specifications extracted from code (\textit{CodeSpec}, from Stage 2) and protocol standards (\textit{DocSpec}, from Stage 3). Using \textit{Differential Analysis}, we identify discrepancies such as mismatched fields or constraints. Each inconsistency is manually examined to determine whether it is a false positive or a true semantic mismatch. Confirmed mismatches are then classified as either parser bugs or issues in the protocol standard.

\section{Evaluation}\label{sec:eval}
Our tool is built on top of AutoGen~\cite{AutoGen}, a multi-agent framework for developing LLM applications.
We use Claude-3.5 Sonnet~\cite{claude} as the LLM API with a temperature of 0 to reduce randomness and enhance reproducibility.
For repository interaction, we build our tool using Tree-sitter~\cite{treesitter}, an AST-based parser to analyze and manipulate source code.
For protocol format extraction (\Cref{subsec:stage2} and \Cref{subsec:stage3}), we use 3D language~\cite{Spec3D} as the DSL and EverParse~\cite{syntaxchecker} as the syntax checker.
We evaluate the effectiveness of our approach by addressing the following research questions:
\vspace{1mm}
\begin{itemize}[leftmargin=0.6cm]
    \item \textbf{RQ1}: How effective are the four stages of \sysname?
    \item \textbf{RQ2}: How well does the LLM extract format specifications from code compared to a baseline?
    \item \textbf{RQ3}: What are the root causes of the discrepancies between the implementation and the RFC? 
    \item \textbf{RQ4}: How much manual effort does \sysname require?

\end{itemize}

\subsection{Dataset}
We evaluate \sysname on the BFD protocol using the C-based implementation from FRRouting~\cite{frr} (stable/8.4), and RFC 5880~\cite{BFD} as the official protocol standard. \sysname is then applied to validate the implementation against the RFC.

\subsection{RQ1: Effectiveness of Each Stage in \sysname}
\subsubsection{Stage 1: Parser Isolation}
We evaluate parser isolation by measuring the precision and recall of the resulting \textit{isolated parsing module}. Precision reflects the correctness of extracted parsing logic, while recall indicates how much of the original logic is preserved. Ground truth is established by manually annotating the original parser, aided by NetLifter~\cite{shi2023lifting}, which uses LLVM-based static analysis to identify parsing-relevant IR instructions.

\vspace{1mm}
\noindent\textbf{Results.}
The \textit{isolated parsing module} achieves 100\% precision and recall, showing that all validation logic is correctly extracted and preserved. This demonstrates \sysname’s ability to isolate complete and accurate parsing logic.
We also assess whether isolation alters parser behavior.
In the BFD case, packet parsing does not depend on runtime context (e.g., socket type or protocol version), so the isolated module maintains the intended input structure. However, for protocols influenced by configuration or negotiation state, isolation may unintentionally broaden or restrict accepted inputs—requiring additional control flow modeling to preserve these semantics.

\subsubsection{Stage 2: Specification Extraction from Code}\label{subsubsec: stage2}
We evaluate \sysname’s accuracy in generating \textit{CodeSpec}, the format specification extracted from the \textit{isolated parsing module}, consisting of field names, types, and constraints. A ground truth is manually constructed from the module’s logic, and precision and recall are used to measure the correctness and completeness of the extracted specification.

\begin{table}[h]
    \centering
    \caption{Precision and Recall for Format Specification Extraction from the \textit{Isolated Parsing Module}}
    \label{tab:format_extraction_code}
    \begin{tabular}{l|ccc}
        \toprule
        \textbf{Metric} & \sysname &\textbf{Precision} & \textbf{Recall} \\
        \midrule
        Field Name  & 11 & 100.0\% & 100.0\% \\
        Field Type  & 11 & 100.0\% & 100.0\% \\
        Field Constraint & 4  & 100.0\% & 100.0\% \\
        \bottomrule
    \end{tabular}
\end{table}

\vspace{1mm}
\noindent\textbf{Results.}
As shown in Table \ref{tab:format_extraction_code}, \sysname correctly extracts all 11 field names, 11 field types, and 4 field constraints with perfect precision and recall. Since the isolated parser fully encapsulates the original parsing logic, the extracted \textit{CodeSpec} is both accurate and complete for this target.

\subsubsection{Stage 3: Specification Extraction from Doc}
We evaluate \textit{DocSpec}, the format extracted from the RFC, by comparing it against a manually constructed ground truth, using precision and recall to assess accuracy and completeness.
\begin{table}[h]
    \centering
    \caption{Precision and Recall for Format Specification Extraction from RFC Document}
    \label{tab:format_extraction_rfc}
    \begin{tabular}{l|ccc}
        \toprule
        \textbf{Metric} & \sysname &\textbf{Precision} & \textbf{Recall} \\
        \midrule
        Field Name  & 37 & 100.0\% & 100.0\% \\
        Field Type  & 37 & 97.3\% & 97.7\% \\
        Field Constraint & 15 & 100.0\% & 58.3\% \\
        \bottomrule
    \end{tabular}
\end{table}

\vspace{1mm}
\noindent\textbf{Results.}
As shown in \Cref{tab:format_extraction_rfc}, \sysname extracts 37 field names, 37 field types, and 15 field constraints from the RFC document.
Field names and types are extracted with near-perfect precision and recall, as they are explicitly defined.
Constraint extraction is more difficult. While the precision remains 100\%, recall drops to 58.3\%, this is because many constraints are implied rather than explicitly stated in the RFC documents.
Despite some missing constraints, the high precision indicates that those extracted constraints are reliable and useful for validation. In particular, if an implementation fails to enforce a constraint specified in \textit{DocSpec}, it is highly likely to indicate a true implementation bug.

\subsubsection{Stage 4: Parser Validation}
We validate the parser by comparing \textit{CodeSpec} and \textit{DocSpec}, focusing on field type and constraint differences, as field names do not affect behavior.
Discrepancies are identified by mismatches between the two specifications, and then traced back to source code and the RFC to determine whether they represent true inconsistencies.

\begin{table}[h]
    \centering
    \caption{Differential Analysis Results: \textit{CodeSpec} vs. \textit{DocSpec}}
    \label{tab:format_discrepancy_analysis}
    \resizebox{\columnwidth}{!}{%
    \begin{tabular}{l|ccc}
        \toprule
        \textbf{Discrepancy Type} & \textbf{Total Inconsis.} & \textbf{True Inconsis.} & \textbf{Extraction Error} \\
        \midrule
        Field Type  & 21  & 21 & 0\\
        Field Constraint  & 15  & 13 & 2\\
        \midrule
        Total                & 36  & 34 & 2\\
        \bottomrule
    \end{tabular}
    }
\end{table}

\vspace{1mm}
\noindent\textbf{Results.}
As shown in \Cref{tab:format_discrepancy_analysis}, \sysname reports 36 discrepancies between \textit{CodeSpec} and \textit{DocSpec}, including 21 type and 15 constraint mismatches. Among the 36 discrepancies, 34 are confirmed as true inconsistencies, and only two are false positives caused by LLM hallucination, yielding a low 5.6\% false positive rate. This demonstrates \sysname's effectiveness in detecting real mismatches between parser implementation and its protocol standard.

\subsection{RQ2: Baseline Comparison}
\noindent\textbf{Setup and Metrics.} 
To assess \sysname's end-to-end effectiveness in extracting format specification from source code (including both parser isolation (Stage 1) and specification extraction (Stage 2)), we compare \textit{CodeSpec} against a manually constructed ground truth based on the full BFD implementation, not just  the \textit{isolated parsing module}.
Precision and recall are used to measure how well \sysname captures field names, types, and constraints.

\noindent\textbf{Results.}
As shown in Table~\ref{tab:format_extraction_code_2}, \sysname achieves 100\% precision and recall, confirming its accuracy and completeness against the full implementation.

\begin{table}[h]
    \centering
    \caption{Precision and Recall for Format Specification Extraction from the Full BFD Implementation}
    \label{tab:format_extraction_code_2}
    \begin{tabular}{l|ccc}
        \toprule
        \textbf{Metric} & \sysname &\textbf{Precision} & \textbf{Recall} \\
        \midrule
        Field Name  & 11 & 100.0\% & 100.0\% \\
        Field Type  & 11 & 100.0\% & 100.0\% \\
        Field Constraint & 4  & 100.0\% & 100.0\% \\
        \bottomrule
    \end{tabular}
\end{table}

\subsection{RQ3: Root Cause of Identified Discrepancies}
\noindent\textbf{Setup and Metrics.} 
To identify the root causes of true discrepancies, we manually analyze each true inconsistency and classify it as either an implementation bug or an RFC issue such as unclear, inaccurate, or missing descriptions.
For implementation bugs, we check existing bug reports to determine whether they are newly discovered. 
For RFC issues, we carefully examine the RFC descriptions to assess their root causes.

\begin{table}[h]
    \centering
    \caption{Root Causes of Identified Inconsistencies}
    \label{tab:discrepancy_classification}
    \resizebox{0.48\textwidth}{!}{
    \begin{tabular}{l|cc}
        \toprule
        \textbf{Discrepancy Type} & \textbf{Implementation Bug} & \textbf{RFC Issue } \\
        \midrule
        Field Type Mismatch   &  21& 0 \\

        Constraint Mismatch   & 11 & 2  \\
        \midrule
        Total & 32 & 2\\
        \bottomrule
    \end{tabular}
    }
\end{table}

\vspace{1mm}
\noindent\textbf{Results.} 
As shown in Table \ref{tab:discrepancy_classification}, 32 of 34 true inconsistencies are implementation bugs, while two stem from RFC issues.
To avoid redundancy in reporting, we group inconsistencies with the same root cause and identify seven unique bugs, five of which are newly detected by \sysname, detailed in \Cref{tab:bugs}. The two RFC-related issues are also previously undocumented and detailed in \Cref{tab:rfcissue}.

\begin{table}[h]
    \centering
    \caption{The Detected Implementation Bugs}
    \label{tab:bugs}
    \resizebox{\columnwidth}{!}{
    \begin{tabular}{clc}
        \toprule
        \textbf{No.} & \textbf{Bug Description} & \textbf{New} \\
        \midrule
        1 & Flag \texttt{M} should always be 0. & \ding{55}\\
        2 & Miss check Authentication Present (A) to handle optional Authentication Section. & \ding{55} \\
        3 & Miss validation for Simple Password Authentication Section format. & \ding{51} \\
        4 & Miss validation for Keyed MD5 Authentication Section format. & \ding{51} \\
        5 & Miss validation for Meticulous Keyed MD5 Authentication Section format. & \ding{51}\\
        6 & Miss validation for Keyed SHA1 Authentication Section Format. & \ding{51}\\
        7 & Miss validation for Meticulous Keyed SHA1 Authentication Section format. & \ding{51} \\
        \bottomrule
    \end{tabular}
    }
\end{table}

\begin{table}[h]
    \centering
    \caption{The Detected RFC Document Issues}
    \label{tab:rfcissue}
    \resizebox{0.48\textwidth}{!}{
    \begin{tabular}{clc}
        \toprule
        \textbf{No.} & \textbf{RFC Document Issues} &\textbf{New} \\
        \midrule
        1 &  Miss explicitly mention that \texttt{Detect Mult} should not be 0. & \ding{51}\\
        2 & Miss explicitly mention that \texttt{Length} should be at least 24.  & \ding{51}\\
        \bottomrule
    \end{tabular}
    }
\end{table}

\subsection{RQ4: Estimation of Manual Effort}
We assess the manual effort involved in each stage of \sysname.
Stage 1 requires manually identifying the entry parsing function, which takes approximately 10 minutes.
Stages 2 and 3 are fully automated and require no human intervention.
Stage 4, which involves comparing CodeSpec and DocSpec and analyzing inconsistencies, is currently manual and takes about 20 minutes.

\section{Discussion}\label{sec:discussion}
This work explores the potential of LLMs to validate network protocol parsers by extracting and comparing protocol format specifications from source code and natural language documents. Rather than providing a fully verified or complete solution, \sysname demonstrates how LLMs can assist in a domain traditionally dominated by static analysis and formal methods.
While promising, \sysname does not provide formal soundness or completeness guarantees. Both \textit{CodeSpec} and \textit{DocSpec} are LLM-generated and may contain inaccuracies, requiring manual inspection to distinguish true inconsistencies from extraction errors.

Given these limitations, \sysname is best viewed as a practical assistant that complements formal techniques.
Its effectiveness must be evaluated empirically. Although we demonstrate its utility on BFD, broader validation across diverse protocols and implementations is needed to assess generality and robustness. Future work includes extending support to parsers written in other programming languages, further automating the validation process, and integrating \sysname into existing protocol testing frameworks.

\section{Related Work}\label{sec:related}
\subsection{Techniques for Protocol Parser Validation}
\noindent\textbf{Model Checking.}
Model checking-based methods~\cite{MusuvathiE04, DiazCRP04, Magic} formally verifies protocol implementations against their specifications. 
However, constructing formal models often requires significant manual effort, limiting its applicability to practical validation tasks.

\vspace{1mm}
\noindent\textbf{Conventional Fuzzing.}
Fuzzing is a widely used approach for testing protocol implementations by generating test cases and executing them to trigger unexpected behavior, particularly crashes. Mutation-based fuzzing, such as AFLNET~\cite{pham2020aflnet}, creates test inputs by mutating existing valid or semi-valid inputs. While it requires little knowledge of the protocol format, its effectiveness relies heavily on the quality of seed inputs and often struggles to reach deep execution paths when parsing conditions are complex. In contrast, generation-based fuzzing, like BooFuzz~\cite{boofuzz}, generates inputs using predefined grammars, improving coverage but requiring additional effort to define these grammars. NetLifter~\cite{shi2023lifting} automates grammar construction through static analysis to extract path conditions, but it still focuses only on low level bugs such as crashes.
However, many protocol implementation bugs are semantic bugs that do not cause crashes but violate protocol specifications, which conventional fuzzing fails to detect.

\vspace{1mm}
\noindent\textbf{Differential Analysis.}
Differential analysis identifies inconsistencies by comparing multiple implementations of the same protocol.
Static tools like ParDiff~\cite{Pardiff} extract state machines from different protocol implementations to detect deviations.
Dynamic differential testing tools, like DPIFuzz~\cite{reen2020dpifuzz} and Prognosis~\cite{ferreira2021prognosis}, generate test cases to execute across multiple implementations and compare their runtime behaviors.
While these approaches can uncover semantic bugs (i.e., silent violations of protocol rules), they rely on the availability of multiple independent implementations, which is not always feasible. 
They also fail to detect bugs shared across all implementations, limiting their effectiveness in certain scenarios.

\subsection{Protocol Format Lifting}
Previous work has focused on extracting protocol formats from either source code or protocol standards to support validation and testing.

\vspace{1mm}
\noindent\textbf{Format Extraction from Source Code.} Tools like Netlifter~\cite{shi2023lifting} and ParDiff~\cite{Pardiff} use static analysis to extract protocol formats from source code.
These methods often struggle with loops, either relying on imprecise heuristics for loop summarization, or bounded loop unrolling, leading to incomplete or inaccurate format extraction.
Additionally, these tools are typically language-specific and require significant effort to support new languages.
\sysname overcomes these limitations by leveraging LLMs for flexible, language-agnostic format extraction. It summarizes loop behavior and validates the extracted logic through test case execution.

\vspace{1mm}
\noindent\textbf{Format Extraction from Protocol Standards.} 
Recent works~\cite{fakhoury20243dgen, chatafl, ParCleanse} highlight the potential of LLMs for extracting format specifications from protocol standards.
3DGen~\cite{fakhoury20243dgen} and ParCleanse~\cite{ParCleanse}  apply LLMs to transform unstructured protocol descriptions into structured formats. In contrast, ChatAFL~\cite{chatafl} avoids feeding RFCs to the model and instead relies on pre-trained models like GPT-3.5, which have been trained on RFCs and are able to answer format-related queries directly.

\sysname combines both source code and document-based format extraction, enabling cross-validation to identify inconsistencies between implementations and their protocol standards. By leveraging LLMs, \sysname offers a flexible, language-agnostic approach to protocol format lifting.

\subsection{Large Language Models for Coding Tasks}
Large Language Models (LLMs) have been widely used for various coding tasks, such as code generation~\cite{starcoder, ding2024cycle, liu2024evaluating, zhu2024deepseek}, program testing~\cite{deng2023large, kang2023large, yang2023whitefox, 10.1145/3643769}, static bug detection~\cite{llm-vul-2, li2024enhancing, DBLP:conf/emnlp/WangZSX024, DBLP:conf/nips/WangZSXX024, guo2025repoaudit, DBLP:journals/corr/abs-2405-17238}, reverse engineering~\cite{xie2024resym, xu2023leveraging, tan2024llm4decompile}, and automated repair~\cite{jiang2023impact, zhang2024autocoderover, llm-vul-1}.
These models leverage extensive code repositories, natural language data, and domain-specific knowledge to tackle tasks that once demanded substantial human expertise~\cite{dubey2024llama, zhu2024deepseek, liu2024deepseek,brown2020language, xie2023impact}.
However, LLM-based approaches also face challenges such as hallucination~\cite{wang2022recode, liu2024exploring} and incapabilities in reasoning about complex program behaviors~\cite{chen2024reasoning, liu2024codemind}. 
To overcome these challenges, recent research has explored techniques such as chain-of-thought prompting~\cite{wei2022chain}, retrieval-augmented generation~\cite{minor2024retrieval}, and the incorporation of symbolic reasoning~\cite{li2024enhancing, DBLP:conf/emnlp/WangZSX024, DBLP:conf/nips/WangZSXX024}, aiming to enhance both the reliability and transparency of model outputs.
\sysname builds on these advancements by leveraging retrieval-augmented-LLMs for protocol format extraction and using symbolic test case generation for refinement, thereby reducing hallucinations and enhancing correctness.
\section{Conclusion}\label{sec:conclusion}
This paper presents \sysname, a multi-agent framework that leverages LLMs to validate network protocol parsers against protocol standards.
By extracting format specifications from both source code and natural language documentation, \sysname enables direct comparison to uncover inconsistencies between implementations and standards.
We evaluate \sysname on the BFD protocol and demonstrate its effectiveness in identifying mismatches with high precision, uncovering seven unique implementations bugs, five of which were previously unknown. While \sysname does not offer formal guarantees, it illustrates the practical utility of LLMs in improving protocol correctness. Our results underscore the potential of LLM-assisted validation as a complement to traditional static analysis and formal methods, offering a more scalable
approach to protocol verification.
\section*{Acknowledgment}
We thank all the anonymous reviewers and our shepherd, Nathan Dautenhahn, for the insightful feedback and guidance.
We are grateful to the Center for AI Safety for providing computational resources. This work was funded in part by the National Science Foundation (NSF) Awards SHF-1901242, SHF-1910300, Proto-OKN 2333736, IIS-2416835, DARPA VSPELLS - HR001120S0058, and ONR N00014-23-1-2081. Any opinions, findings and conclusions or recommendations expressed in this material are those of the authors and do not necessarily reflect the views of the sponsors.

\bibliographystyle{IEEEtran}
\bibliography{ref}

\begin{thebibliography}{10}
\providecommand{\url}[1]{#1}
\csname url@samestyle\endcsname
\providecommand{\newblock}{\relax}
\providecommand{\bibinfo}[2]{#2}
\providecommand{\BIBentrySTDinterwordspacing}{\spaceskip=0pt\relax}
\providecommand{\BIBentryALTinterwordstretchfactor}{4}
\providecommand{\BIBentryALTinterwordspacing}{\spaceskip=\fontdimen2\font plus
\BIBentryALTinterwordstretchfactor\fontdimen3\font minus \fontdimen4\font\relax}
\providecommand{\BIBforeignlanguage}[2]{{%
\expandafter\ifx\csname l@#1\endcsname\relax
\typeout{** WARNING: IEEEtran.bst: No hyphenation pattern has been}%
\typeout{** loaded for the language `#1'. Using the pattern for}%
\typeout{** the default language instead.}%
\else
\language=\csname l@#1\endcsname
\fi
#2}}
\providecommand{\BIBdecl}{\relax}
\BIBdecl

\bibitem{TLS}
``The transport layer security (tls) protocol,'' \url{https://datatracker.ietf.org/doc/html/rfc5246}.

\bibitem{FTP}
``File transfer protocol (ftp),'' \url{https://datatracker.ietf.org/doc/html/rfc959}.

\bibitem{IP}
``Internet protocol (ip),'' \url{https://datatracker.ietf.org/doc/html/rfc791}.

\bibitem{heartbleed}
Heartbleed, ``The heartbleed bug,'' \url{https://heartbleed.com}, 2020.

\bibitem{OpenSSL}
``Openssl,'' \url{https://www.openssl.org}.

\bibitem{CVE-2021-41773}
``Cve-2021-41773,'' \url{https://cve.mitre.org/cgi-bin/cvename.cgi?name=CVE-2021-41773}.

\bibitem{MusuvathiE04}
M.~Musuvathi and D.~R. Engler, ``Model checking large network protocol implementations,'' in \emph{{NSDI}}.\hskip 1em plus 0.5em minus 0.4em\relax {USENIX}, 2004, pp. 155--168.

\bibitem{DiazCRP04}
G.~D{\'{\i}}az, F.~Cuartero, V.~V. Ruiz, and F.~L. Pelayo, ``Automatic verification of the {TLS} handshake protocol,'' in \emph{{SAC}}.\hskip 1em plus 0.5em minus 0.4em\relax {ACM}, 2004, pp. 789--794.

\bibitem{Magic}
S.~Chaki, E.~M. Clarke, A.~Groce, S.~Jha, and H.~Veith, ``Modular verification of software components in {C},'' in \emph{{ICSE}}.\hskip 1em plus 0.5em minus 0.4em\relax {IEEE} Computer Society, 2003, pp. 385--395.

\bibitem{pham2020aflnet}
V.-T. Pham, M.~B{\"o}hme, and A.~Roychoudhury, ``Aflnet: a greybox fuzzer for network protocols,'' in \emph{2020 IEEE 13th International Conference on Software Testing, Validation and Verification (ICST)}.\hskip 1em plus 0.5em minus 0.4em\relax IEEE, 2020, pp. 460--465.

\bibitem{shi2023lifting}
Q.~Shi, J.~Shao, Y.~Ye, M.~Zheng, and X.~Zhang, ``Lifting network protocol implementation to precise format specification with security applications,'' in \emph{Proceedings of the 2023 ACM SIGSAC Conference on Computer and Communications Security}, ser. CCS '23.\hskip 1em plus 0.5em minus 0.4em\relax ACM, 2023, p. 1287–1301.

\bibitem{Pardiff}
M.~Zheng, Q.~Shi, X.~Liu, X.~Xu, L.~Yu, C.~Liu, G.~Wei, and X.~Zhang, ``Pardiff: Practical static differential analysis of network protocol parsers,'' in \emph{Proc. ACM Program. Lang.}, ser. OOPSLA '24.\hskip 1em plus 0.5em minus 0.4em\relax ACM, 2024, pp. 1208--1234.

\bibitem{KIT}
\BIBentryALTinterwordspacing
C.~Liu, S.~Gong, and P.~Fonseca, ``Kit: Testing os-level virtualization for functional interference bugs,'' in \emph{Proceedings of the 28th ACM International Conference on Architectural Support for Programming Languages and Operating Systems, Volume 2}, ser. ASPLOS 2023.\hskip 1em plus 0.5em minus 0.4em\relax New York, NY, USA: Association for Computing Machinery, 2023, p. 427–441. [Online]. Available: \url{https://doi.org/10.1145/3575693.3575731}
\BIBentrySTDinterwordspacing

\bibitem{9678600}
M.~Zheng, J.~Yang, M.~Wen, H.~Zhu, Y.~Liu, and H.~Jin, ``Why do developers remove lambda expressions in java?'' in \emph{2021 36th IEEE/ACM International Conference on Automated Software Engineering (ASE)}, 2021, pp. 67--78.

\bibitem{ParCleanse}
M.~Zheng, D.~Xie, Q.~Shi, C.~Wang, and X.~Zhang, ``Validating network protocol parsers with traceable rfc document interpretation,'' in \emph{Proceedings of the 34th ACM SIGSOFT International Symposium on Software Testing and Analysis}, ser. ISSTA 2025.

\bibitem{reen2020dpifuzz}
G.~S. Reen and C.~Rossow, ``Dpifuzz: a differential fuzzing framework to detect dpi elusion strategies for quic,'' in \emph{Proceedings of the 36th Annual Computer Security Applications Conference}, ser. ACSAC '20.\hskip 1em plus 0.5em minus 0.4em\relax ACM, 2020, pp. 332--344.

\bibitem{fakhoury20243dgen}
S.~Fakhoury, M.~Kuppe, S.~K. Lahiri, T.~Ramananandro, and N.~Swamy, ``3dgen: Ai-assisted generation of provably correct binary format parsers,'' \emph{arXiv preprint arXiv:2404.10362}, 2024.

\bibitem{treesitter}
``Tree-sitter,'' \url{https://tree-sitter.github.io/tree-sitter/}.

\bibitem{AutoGen}
\BIBentryALTinterwordspacing
Q.~Wu, G.~Bansal, J.~Zhang, Y.~Wu, B.~Li, E.~Zhu, L.~Jiang, X.~Zhang, S.~Zhang, J.~Liu, A.~H. Awadallah, R.~W. White, D.~Burger, and C.~Wang, ``Autogen: Enabling next-gen {LLM} applications via multi-agent conversation,'' in \emph{ICLR 2024 Workshop on Large Language Model (LLM) Agents}, 2024. [Online]. Available: \url{https://openreview.net/forum?id=uAjxFFing2}
\BIBentrySTDinterwordspacing

\bibitem{claude}
``Claude 3.5 sonnet,'' \url{https://www.anthropic.com/claude/sonnet}.

\bibitem{Spec3D}
``3d: Dependent data descriptions for verified validation,'' \url{https://project-everest.github.io/everparse/3d.html}.

\bibitem{syntaxchecker}
``Everparse,'' https://project-everest.github.io/everparse/3d-lang.html.

\bibitem{frr}
F.~community, ``The frrouting protocol suite,'' \url{https://github.com/FRRouting/frr}, 2024.

\bibitem{BFD}
``Bidirectional forwarding detection (bfd),'' \url{https://datatracker.ietf.org/doc/html/rfc5880}.

\bibitem{boofuzz}
J.~Pereyda, ``Boofuzz,'' \url{https://github.com/jtpereyda/boofuzz}, 2023.

\bibitem{ferreira2021prognosis}
T.~Ferreira, H.~Brewton, L.~D'Antoni, and A.~Silva, ``Prognosis: closed-box analysis of network protocol implementations,'' in \emph{Proceedings of the 2021 ACM SIGCOMM 2021 Conference}, 2021, pp. 762--774.

\bibitem{chatafl}
R.~Meng, M.~Mirchev, M.~B{\"o}hme, and A.~Roychoudhury, ``Large language model guided protocol fuzzing,'' in \emph{Proceedings of the 31st Annual Network and Distributed System Security Symposium (NDSS)}, ser. NDSS '24.\hskip 1em plus 0.5em minus 0.4em\relax The Internet Society, 2024.

\bibitem{starcoder}
R.~Li, L.~B. Allal, Y.~Zi, N.~Muennighoff, D.~Kocetkov, C.~Mou, M.~Marone, C.~Akiki, J.~Li, J.~Chim \emph{et~al.}, ``Starcoder: may the source be with you!'' \emph{arXiv preprint arXiv:2305.06161}, 2023.

\bibitem{ding2024cycle}
Y.~Ding, M.~J. Min, G.~Kaiser, and B.~Ray, ``Cycle: Learning to self-refine the code generation,'' \emph{Proceedings of the ACM on Programming Languages}, vol.~8, no. OOPSLA1, pp. 392--418, 2024.

\bibitem{liu2024evaluating}
\BIBentryALTinterwordspacing
J.~Liu, S.~Xie, J.~Wang, Y.~Wei, Y.~Ding, and L.~Zhang, ``Evaluating language models for efficient code generation,'' in \emph{First Conference on Language Modeling}, 2024. [Online]. Available: \url{https://openreview.net/forum?id=IBCBMeAhmC}
\BIBentrySTDinterwordspacing

\bibitem{zhu2024deepseek}
Q.~Zhu, D.~Guo, Z.~Shao, D.~Yang, P.~Wang, R.~Xu, Y.~Wu, Y.~Li, H.~Gao, S.~Ma \emph{et~al.}, ``Deepseek-coder-v2: Breaking the barrier of closed-source models in code intelligence,'' \emph{arXiv preprint arXiv:2406.11931}, 2024.

\bibitem{deng2023large}
Y.~Deng, C.~S. Xia, H.~Peng, C.~Yang, and L.~Zhang, ``Large language models are zero-shot fuzzers: Fuzzing deep-learning libraries via large language models,'' in \emph{Proceedings of the 32nd ACM SIGSOFT international symposium on software testing and analysis}, 2023, pp. 423--435.

\bibitem{kang2023large}
S.~Kang, J.~Yoon, and S.~Yoo, ``Large language models are few-shot testers: Exploring llm-based general bug reproduction,'' in \emph{2023 IEEE/ACM 45th International Conference on Software Engineering (ICSE)}.\hskip 1em plus 0.5em minus 0.4em\relax IEEE, 2023, pp. 2312--2323.

\bibitem{yang2023whitefox}
C.~Yang, Y.~Deng, R.~Lu, J.~Yao, J.~Liu, R.~Jabbarvand, and L.~Zhang, ``Whitefox: White-box compiler fuzzing empowered by large language models,'' \emph{arXiv preprint arXiv:2310.15991}, 2023.

\bibitem{10.1145/3643769}
\BIBentryALTinterwordspacing
G.~Ryan, S.~Jain, M.~Shang, S.~Wang, X.~Ma, M.~K. Ramanathan, and B.~Ray, ``Code-aware prompting: A study of coverage-guided test generation in regression setting using llm,'' vol.~1, no. FSE, Jul. 2024. [Online]. Available: \url{https://doi.org/10.1145/3643769}
\BIBentrySTDinterwordspacing

\bibitem{llm-vul-2}
B.~Steenhoek, M.~M. Rahman, R.~Jiles, and W.~Le, ``An empirical study of deep learning models for vulnerability detection,'' in \emph{2023 IEEE/ACM 45th International Conference on Software Engineering (ICSE)}, 2023, pp. 2237--2248.

\bibitem{li2024enhancing}
H.~Li, Y.~Hao, Y.~Zhai, and Z.~Qian, ``Enhancing static analysis for practical bug detection: An llm-integrated approach,'' \emph{Proceedings of the ACM on Programming Languages}, vol.~8, no. OOPSLA1, pp. 474--499, 2024.

\bibitem{DBLP:conf/emnlp/WangZSX024}
\BIBentryALTinterwordspacing
C.~Wang, W.~Zhang, Z.~Su, X.~Xu, and X.~Zhang, ``Sanitizing large language models in bug detection with data-flow,'' in \emph{Findings of the Association for Computational Linguistics: {EMNLP} 2024, Miami, Florida, USA, November 12-16, 2024}, Y.~Al{-}Onaizan, M.~Bansal, and Y.~Chen, Eds.\hskip 1em plus 0.5em minus 0.4em\relax Association for Computational Linguistics, 2024, pp. 3790--3805. [Online]. Available: \url{https://aclanthology.org/2024.findings-emnlp.217}
\BIBentrySTDinterwordspacing

\bibitem{DBLP:conf/nips/WangZSXX024}
\BIBentryALTinterwordspacing
C.~Wang, W.~Zhang, Z.~Su, X.~Xu, X.~Xie, and X.~Zhang, ``{LLMDFA:} analyzing dataflow in code with large language models,'' in \emph{Advances in Neural Information Processing Systems 38: Annual Conference on Neural Information Processing Systems 2024, NeurIPS 2024, Vancouver, BC, Canada, December 10 - 15, 2024}, A.~Globersons, L.~Mackey, D.~Belgrave, A.~Fan, U.~Paquet, J.~M. Tomczak, and C.~Zhang, Eds., 2024. [Online]. Available: \url{http://papers.nips.cc/paper\_files/paper/2024/hash/ed9dcde1eb9c597f68c1d375bbecf3fc-Abstract-Conference.html}
\BIBentrySTDinterwordspacing

\bibitem{guo2025repoaudit}
J.~Guo, C.~Wang, X.~Xu, Z.~Su, and X.~Zhang, ``Repoaudit: An autonomous llm-agent for repository-level code auditing,'' \emph{arXiv preprint arXiv:2501.18160}, 2025.

\bibitem{DBLP:journals/corr/abs-2405-17238}
\BIBentryALTinterwordspacing
Z.~Li, S.~Dutta, and M.~Naik, ``Llm-assisted static analysis for detecting security vulnerabilities,'' \emph{CoRR}, vol. abs/2405.17238, 2024. [Online]. Available: \url{https://doi.org/10.48550/arXiv.2405.17238}
\BIBentrySTDinterwordspacing

\bibitem{xie2024resym}
D.~Xie, Z.~Zhang, N.~Jiang, X.~Xu, L.~Tan, and X.~Zhang, ``Resym: Harnessing llms to recover variable and data structure symbols from stripped binaries,'' in \emph{Proceedings of the 2024 on ACM SIGSAC Conference on Computer and Communications Security}, 2024, pp. 4554--4568.

\bibitem{xu2023leveraging}
X.~Xu, Z.~Zhang, Z.~Su, Z.~Huang, S.~Feng, Y.~Ye, N.~Jiang, D.~Xie, S.~Cheng, L.~Tan \emph{et~al.}, ``Leveraging generative models to recover variable names from stripped binary,'' \emph{arXiv preprint arXiv:2306.02546}, 2023.

\bibitem{tan2024llm4decompile}
H.~Tan, Q.~Luo, J.~Li, and Y.~Zhang, ``Llm4decompile: Decompiling binary code with large language models,'' \emph{arXiv preprint arXiv:2403.05286}, 2024.

\bibitem{jiang2023impact}
N.~Jiang, K.~Liu, T.~Lutellier, and L.~Tan, ``Impact of code language models on automated program repair,'' in \emph{2023 IEEE/ACM 45th International Conference on Software Engineering (ICSE)}.\hskip 1em plus 0.5em minus 0.4em\relax IEEE, 2023, pp. 1430--1442.

\bibitem{zhang2024autocoderover}
Y.~Zhang, H.~Ruan, Z.~Fan, and A.~Roychoudhury, ``Autocoderover: Autonomous program improvement,'' in \emph{Proceedings of the 33rd ACM SIGSOFT International Symposium on Software Testing and Analysis}, 2024, pp. 1592--1604.

\bibitem{llm-vul-1}
\BIBentryALTinterwordspacing
Y.~Wu, N.~Jiang, H.~V. Pham, T.~Lutellier, J.~Davis, L.~Tan, P.~Babkin, and S.~Shah, ``How effective are neural networks for fixing security vulnerabilities,'' in \emph{Proceedings of the 32nd ACM SIGSOFT International Symposium on Software Testing and Analysis}, ser. ISSTA 2023.\hskip 1em plus 0.5em minus 0.4em\relax New York, NY, USA: Association for Computing Machinery, 2023, p. 1282–1294. [Online]. Available: \url{https://doi.org/10.1145/3597926.3598135}
\BIBentrySTDinterwordspacing

\bibitem{dubey2024llama}
A.~Dubey, A.~Jauhri, A.~Pandey, A.~Kadian, A.~Al-Dahle, A.~Letman, A.~Mathur, A.~Schelten, A.~Yang, A.~Fan \emph{et~al.}, ``The llama 3 herd of models,'' \emph{arXiv preprint arXiv:2407.21783}, 2024.

\bibitem{liu2024deepseek}
A.~Liu, B.~Feng, B.~Xue, B.~Wang, B.~Wu, C.~Lu, C.~Zhao, C.~Deng, C.~Zhang, C.~Ruan \emph{et~al.}, ``Deepseek-v3 technical report,'' \emph{arXiv preprint arXiv:2412.19437}, 2024.

\bibitem{brown2020language}
T.~Brown, B.~Mann, N.~Ryder, M.~Subbiah, J.~D. Kaplan, P.~Dhariwal, A.~Neelakantan, P.~Shyam, G.~Sastry, A.~Askell \emph{et~al.}, ``Language models are few-shot learners,'' \emph{Advances in neural information processing systems}, vol.~33, pp. 1877--1901, 2020.

\bibitem{xie2023impact}
D.~Xie, B.~Yoo, N.~Jiang, M.~Kim, L.~Tan, X.~Zhang, and J.~S. Lee, ``Impact of large language models on generating software specifications,'' \emph{arXiv preprint arXiv:2306.03324}, 2023.

\bibitem{wang2022recode}
S.~Wang, Z.~Li, H.~Qian, C.~Yang, Z.~Wang, M.~Shang, V.~Kumar, S.~Tan, B.~Ray, P.~Bhatia \emph{et~al.}, ``Recode: Robustness evaluation of code generation models,'' \emph{arXiv preprint arXiv:2212.10264}, 2022.

\bibitem{liu2024exploring}
F.~Liu, Y.~Liu, L.~Shi, H.~Huang, R.~Wang, Z.~Yang, L.~Zhang, Z.~Li, and Y.~Ma, ``Exploring and evaluating hallucinations in llm-powered code generation,'' \emph{arXiv preprint arXiv:2404.00971}, 2024.

\bibitem{chen2024reasoning}
J.~Chen, Z.~Pan, X.~Hu, Z.~Li, G.~Li, and X.~Xia, ``Reasoning runtime behavior of a program with llm: How far are we?'' \emph{arXiv preprint cs.SE/2403.16437}, 2024.

\bibitem{liu2024codemind}
C.~Liu, S.~D. Zhang, A.~R. Ibrahimzada, and R.~Jabbarvand, ``Codemind: A framework to challenge large language models for code reasoning,'' \emph{arXiv preprint arXiv:2402.09664}, 2024.

\bibitem{wei2022chain}
J.~Wei, X.~Wang, D.~Schuurmans, M.~Bosma, F.~Xia, E.~Chi, Q.~V. Le, D.~Zhou \emph{et~al.}, ``Chain-of-thought prompting elicits reasoning in large language models,'' \emph{Advances in neural information processing systems}, vol.~35, pp. 24\,824--24\,837, 2022.

\bibitem{minor2024retrieval}
M.~Minor and E.~Kaucher, ``Retrieval augmented generation with llms for explaining business process models,'' in \emph{International Conference on Case-Based Reasoning}.\hskip 1em plus 0.5em minus 0.4em\relax Springer, 2024, pp. 175--190.

\end{thebibliography}

\end{document}